\documentclass[english,superscriptaddress, reprint]{revtex4-1}
\usepackage[LGR,T1]{fontenc}
\usepackage[latin9]{inputenc}
\usepackage{color}
\usepackage{amsmath}
\usepackage{amssymb}
\usepackage{graphicx}
\usepackage{subscript}

\makeatletter

\DeclareRobustCommand{\greektext}{%
  \fontencoding{LGR}\selectfont\def\encodingdefault{LGR}}
\DeclareRobustCommand{\textgreek}[1]{\leavevmode{\greektext #1}}
\ProvideTextCommand{\~}{LGR}[1]{\char126#1}

\makeatother

\usepackage{babel}
\begin{document}

\title{Strong Nonlinear Coupling due to Induced Photon Interaction on a
Si$_{3}$N$_{4}$ Chip}

\author{Sven Ramelow}
\email{Corresponding Author: sven.ramelow@physik.hu-berlin.de}

\affiliation{Institute for Physics, Humboldt University Berlin, Germany}

\author{Alessandro Farsi}
\email{Equally contrubuting author}

\affiliation{Dept. of Appl. Physics and Appl. Math, Columbia University, New York,
USA}

\author{Zachary Vernon }

\affiliation{Dept. of Physics, University of Toronto, Toronto, Canada}

\author{Stephane Clemmen}

\affiliation{Dept. of Inf. Tech., Ghent University, Ghent, Belgium}

\author{Xingchen Ji}

\affiliation{Dept. of Electrical Engineering, Columbia University, New York, USA}

\author{John E. Sipe}

\affiliation{Dept. of Physics, University of Toronto, Toronto, Canada}

\author{Marco Liscidini}

\affiliation{Università degli Studi di Pavia, Pavia, Italy}

\author{Michal Lipson}

\affiliation{Dept. of Electrical Engineering, Columbia University, New York, USA}

\author{Alexander L. Gaeta}

\affiliation{Dept. of Appl. Physics and Appl. Math, Columbia University, New York,
USA}

\maketitle
\textbf{Second-order optical processes lead to a host of applications
in classical and quantum optics. With the enhancement of parametric
interactions that arise due to light confinement, on-chip implementations
promise very-large-scale photonic integration. But as yet there is
no route to a device that acts at the single photon level. Here we
exploit the \textgreek{q}$^{(3)}$ nonlinear response of a Si$_{3}\mathbf{N}_{4}$
microring resonator to induce a large effective \textgreek{q}$^{(2)}$.
Effective second-order upconversion (ESUP) of a seed to an idler can
be achieved with 74,000 \%/W efficiency, indicating that single photon
nonlinearity is within reach of current technology. Moreover, we show
a nonlinear coupling rate of seed and idler larger than the energy
dissipation rate in the resonator, indicating a strong coupling regime.
Consequently we observe a Rabi-like splitting, for which we provide
a detailed theoretical description. This yields new insight into the
dynamics of ultrastrong effective nonlinear interactions in microresonators,
and access to novel phenomena and applications in classical and quantum
nonlinear optics.}

Current strategies for implementing nonlinearity at the single photon
level \citep{imamoglu_strongly_1997,liew_single_2010,ferretti_single-photon_2012}
have to make use of very complex experimental arrangements, such as
high-Q photonic crystal cavities and quantum dots, which are difficult
to implement using conventional photonic technologies. Nonetheless
they are being actively pursued, in part because of the common assumption
that for purely parametric nonlinear optical processes \textendash{}
that is, those that do not involve the excitation of a resonance in
the response of a material medium \textendash{} the intrinsic material
nonlinearities would be far too weak to allow for efficient nonlinear
interactions between single photons. 

This assumption was challenged by Langford et al. \citep{langford_efficient_2011},
who argued that the use of a strong pump for one of the fields of
a four-wave mixing (FWM) process in a $\chi^{(3)}$-nonlinear medium
would lead to an effective $\chi^{(2)}$ nonlinearity for the remaining
three fields. Not directly limited by the intrinsic material properties,
this effective $\chi^{(2)}$ interaction can in principle be made
arbitrarily large by increasing the pump power. While theoretically
this would allow for an induced photon interaction efficient at the
single-photon level \citep{langford_efficient_2011}, to date such
an effective $\chi^{(2)}$ has experimentally only been demonstrated
at relatively low efficiencies using a photonic crystal fiber \citep{langford_efficient_2011},
a chalcogenide nanofiber \citep{meyer-scott_power-efficient_2015}
and a resonant $\chi^{(3)}$-nonlinearity in warm Rb-vapor \citep{donvalkar_quantum_2016}.
Yet such an approach is also compatible with material platforms and
light confinement strategies characterized by strong field enhancements
and exceptionally large Q factors, which offer the promise of strong
effective second-order photon interactions. We shall see below that
this promise can be fulfilled. 

\begin{figure}
\begin{centering}
\includegraphics[width=0.95\columnwidth]{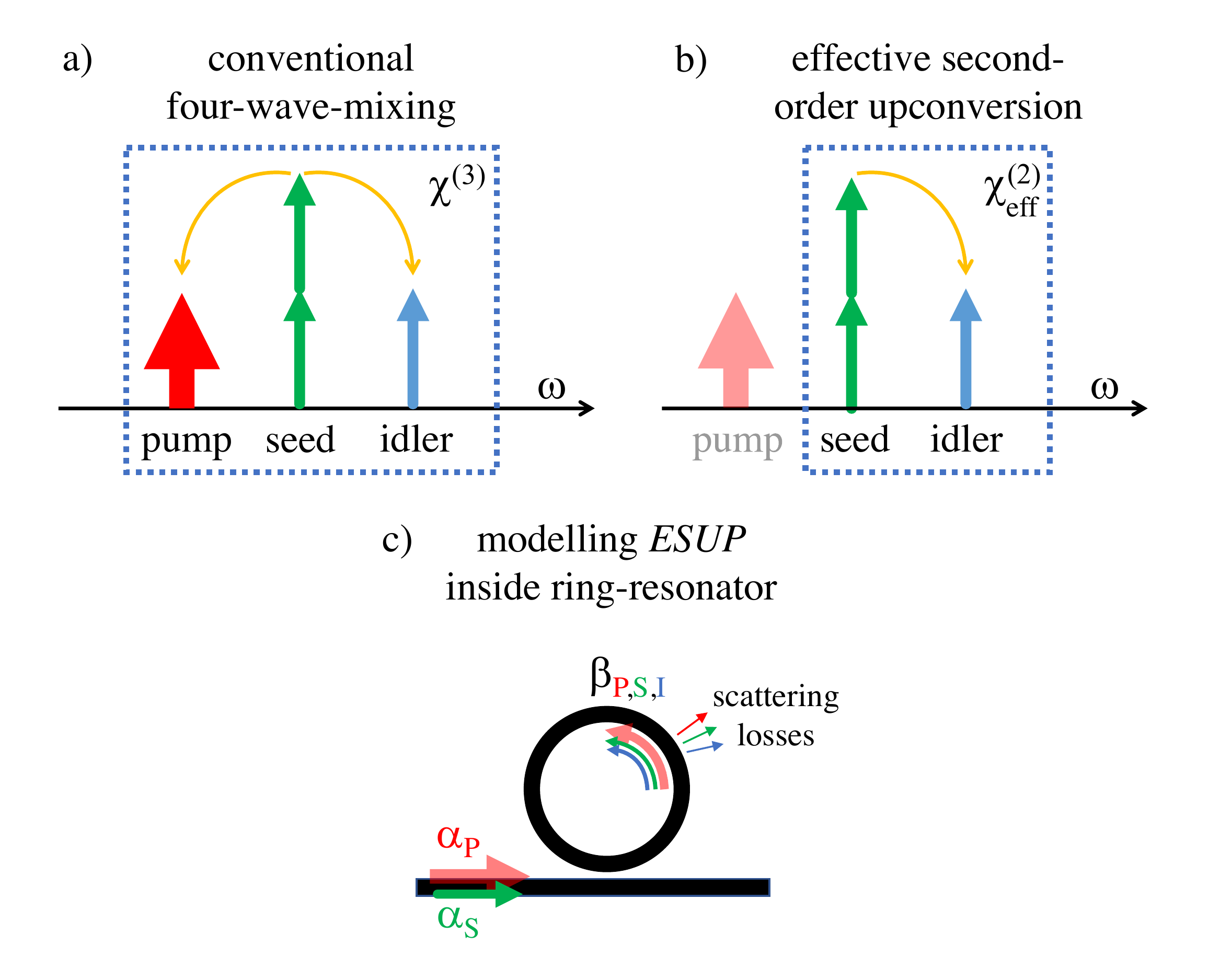}
\par\end{centering}
\caption{\label{fig:energy-cartoon} ESUP conversion: A FWM process driven
by a strong field at ${\color{red}{\normalcolor {\color{red}}\omega_{P}}}$
(a) can be viewed as an effective second-order nonlinear interaction
leading to an up conversion of the seed field. (b). Sketch of the
FWM process in a single-bus waveguide ring resonator (c).}
\end{figure}

In any such scenario, the nonlinear processes that can result are
richer than those occurring in materials with a natural $\chi^{(2)}$,
but are analogous to them. In particular, it is possible to produce
an effective second-order upconversion (ESUP) process (see Fig. \ref{fig:energy-cartoon})
in which two photons are converted to one photon at a higher photon
energy, but with much less than double the original photon energy
that would characterize the analogous process of second harmonic generation
(SHG). To measure the strength of ESUP we employ the standard figure-of-merit
used for quantifying SHG, a normalized efficiency quoted in \%/W.
While ultra-high Q, mm-sized resonators made from lithium niobate
(LN) can reach 300,000 \%/W \citep{lin_phase-matched_2016,furst_naturally_2010},
current state-of-the-art values for integrated devices are on the
order of few thousand \%/mW, such as the 5400 \%/W achieved in LN
waveguides \citep{parameswaran_highly_2002}, and the recently reported
2,500 \%/W in AlN micro-ring resonators \citep{guo_aluminum_2017}.

Here we utilize a high-Q Si\textsubscript{3}N\textsubscript{4} microresonator
to generate a strong effective $\chi^{(2)}$. With a pump field at
$1590$ nm, we induce ESUP of two seed photons at $1400$ nm to an
idler photon at $1260$ nm. All photon energies are at microring resonances,
strongly enhancing the conversion. Changing the seed power at $1400$
nm over 5 orders of magnitude, and measuring the resulting idler power,
we verify the quadratic scaling expected of a second-order process,
and measure a normalized efficiency of 74,000\%/W. Intriguingly, at
higher seed powers of only a few hundreds of microwatts at $1400$
nm, where our total conversion efficiency is saturating at around
5\%, the nonlinear coupling rate in the resonator is comparable with
resonator damping rate. This fulfills a strong coupling condition
between the nonlinearly interacting modes, and consequently there
occurs a Rabi-like splitting of the modes, which we observe and study
both experimentally and theoretically. 

For the theoretical description, we consider the resonant nonlinear
interaction of the three fields schematically indicated in Fig. \ref{fig:energy-cartoon}a:
There is a \textit{strong} pump at $\omega_{P}$, a \textit{weak}
seed at $\omega_{S}$, and an idler generated at $\omega_{I}=2\omega_{S}-\omega_{P}$.
We focus on the effective second-order nonlinear interaction between
the seed and the idler (see \ref{fig:energy-cartoon}b), which can
be described by the equations 

\begin{eqnarray}
\frac{\mathrm{d\beta_{S}}}{dt} & = & -\bar{\Gamma}_{S}\beta_{S}+2i\Lambda_{2}(t)\beta_{S}^{*}\beta_{I}-i\gamma_{S}^{*}\alpha_{S}e^{-i\Delta_{S}t}\label{eq:pw1}\\
\frac{\mathrm{d\beta_{I}}}{dt} & = & -\bar{\Gamma}_{I}\beta_{I}+i\Lambda_{2}^{*}(t)\beta_{S}^{2}\label{eq:pw2}
\end{eqnarray}
where $\beta_{S,I}$ are the slowly varying seed and idler fields
amplitudes in the resonator, $\bar{\Gamma}_{S,I}$ are the total resonator
damping rates at $\omega_{S,I},$ $\alpha_{S}$ the amplitude of the
seed driving field in the bus waveguide, $\Delta_{S}$ is the detuning
away from resonance, and $\gamma_{S}$ characterizes the coupling
between the bus waveguide and the ring (see Fig. \ref{fig:energy-cartoon}b).
The coefficient $\Lambda_{2}(t)$, which describes the strength of
the effective $\chi^{(2)}$, is given by $\Lambda_{2}(t)=\Lambda_{3}\beta_{P}(t)$,
with $\Lambda_{3}\approx\hbar\omega v_{g}^{2}\gamma_{NL}/(2\pi R)$
the third order coupling coefficient, where $\gamma_{NL}$ is the
usual nonlinear parameter, $v_{g}$ the group velocity in the ring
at $\omega_{P}$, $R$ the ring radius, and $\omega^{2}=\sqrt{\omega_{P}\omega_{S}^{2}\omega_{I}}$;
$\beta_{P}(t)$ is the pump amplitude in the ring. We assume that
$\beta_{P}(t)$ is undepleted and that the effects of self- and cross-phase
modulation are taken into account as corrections to the resonant frequencies
$\omega_{S,I,P}$, which is valid for slowly varying field envelope
functions; we restrict our analysis to this regime. See Methods for
details.

The use of a high-Q resonator gives rise to dramatic enhancements
of both the circulating pump power available and the nonlinear interaction
between the three fields \citep{bogaerts_silicon_2012,turner_ultra-low_2008,ferrera_low_2009,gondarenko_cmos-compatible_2010}.
The $\mathrm{Si}{}_{3}\textrm{N}_{4}$ material was chosen to avoid
two-photon absorption, which typically plagues silicon systems. This
enables very large unit power ESUP conversion efficiencies, $\eta_{ESUP}$,
to be achieved. This figure of merit can be theoretically calculated
as 

\begin{align}
 & \eta_{ESUP}=\frac{64R^{2}\gamma_{NL}^{2}(\Delta\lambda_{FSR})^{4}}{\pi^{2}\lambda^{4}}P_{P}^{in}Q^{4}\left(\frac{Q}{Q^{ext}}\right)^{4}\label{eq:eta_sup}
\end{align}
in steady state, where $\Delta\lambda_{FSR}$ is the wavelength free
spectral range, $P_{P}^{in}$ is the pump input power in the bus waveguide,
$Q$ is the loaded quality factor and $Q^{ext}$ the extrinsic quality
factor due to the coupling of the resonator to the bus waveguide.
Corrections to this simplified formula, including the fact that the
quality factors and wavelengths $\lambda$ are different at different
frequencies, are described in the Methods. Importantly, the scaling
of $\eta_{ESUP}$ with $Q^{4}$ is one power larger than its $\chi^{(2)}$analogue
- the normailzed SHG efficiency, which in a resonator only scales
with $Q^{3}$. 

\begin{figure}
\begin{centering}
\includegraphics{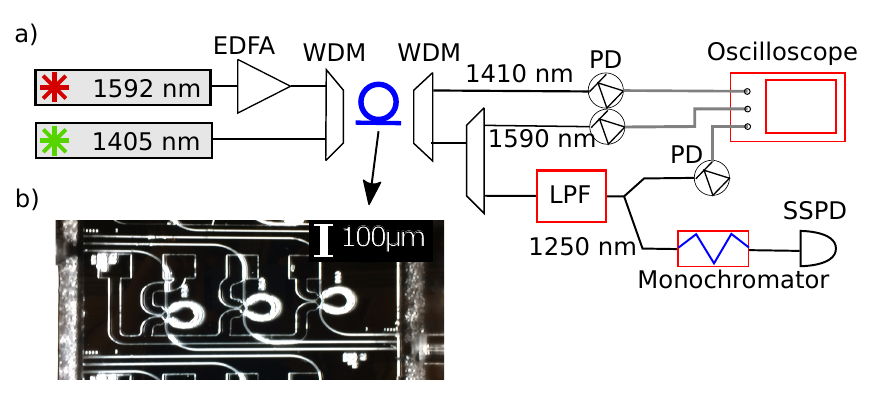}
\par\end{centering}
\caption{\label{fig:schematics}a) Schematics of the experiment. EDFA: erbium
doped fiber amplifier, WDM: wave division multiplexing, LPF: low pass
filter, PD: phododiode, SSPD superconducting single-photon detectors.
b) Microscope image of the microresonator chip and the lensed fiber
for input/output coupling}
\end{figure}

We set out to investigate the ESUP experimentally. The chip-integrated
microring resonators utilized in this work were made from a high quality
Si\textsubscript{3}N\textsubscript{4} film on an SiO\textsubscript{\textcolor{red}{2}}
substrate, grown via low-pressure chemical vapor deposition and electron
beam lithography (see \citep{luke_overcoming_2013} for details).
The waveguide dimensions (Fig.\ref{fig:schematics}b) of $730$ nm
x $910$ nm (height x width) are carefully chosen: With a zero group
velocity dispersion (GVD) at $1410$ nm, the waveguide exhibits phase-matching
between a strong pump at $1590$ nm, the seed field at $1400$ nm,
and the idler field at 1260 nm. Moreover, the large normal GVD at
$\omega_{P}$ strongly inhibits any undesired parametric process (e.g.,
spontaneous FWM) from the strong pump . The microring resonator (radius
$45$ $\mu m$) with a free spectral range of $500$ GHz is coupled
to the bus waveguide with a gap of $450$ nm, and exhibits loaded
Q-factors of 250,000 for the pump (over-coupled), 1.2 million for
the seed (close to critical coupling), and about 3.5 million for the
idler (under-coupled). Using these parameters we theoretically expect
a unit power ESUP efficiency of 87,000 $\pm$ 17,000 \%/W, where the
error interval is estimated from the uncertainty on the experimental
parameters (e.g. coupled power, quality factors, and propagation losses).
This corresponds to a $\chi_{\mathrm{eff}}^{(2)}\simeq1$ pm/V (see
Methods section), which is comparable in magnitude to the
second-order nonlinear susceptibility used in LN WGM-resonators. And
although this is two orders of magnitude smaller than the nonlinear
susceptibility of semiconductors such as GaAs or GaN, the upconversion
efficiency here is among the highest ever reported.

Our measurement scheme is shown in Fig. \ref{fig:schematics}a). Pump
and seedignal light are generated by tunable external cavity lasers.
The pump is amplified with a high power EDFA to mW and combined with
the seed with wavelength division multiplexers (WDMs). They are injected
into the chip (and collected) via lensed fibers with a total insertion
loss of 8.5 dB, resulting in $85$ mW of on-chip pump power at 1590
nm. The output is separated into the three wavelength bands by cascaded
WDMs to be monitored separately on InGaAs photodiodes. The idler output
at 1260 nm is further strongly filtered by a low-pass filter yielding
total losses of $6.3$ dB, including the WDMs. When measuring very
low intensities, we optionally filter the idler with a grating monochromator
($0.55$ nm filter bandwidth), and then detect with superconducting
single-photon detectors. We enclose the chip and stabilize the temperature
to reduce thermal and mechanical effects.

For our measurements of the idler output power at $1260$ nm as a
function of the $1400$ nm seed wavelength, first the $1590$ pump
wavelength is tuned into resonance from the short wavelength side
to follow the thermal red-shift and achieve thermal locking \citep{carmon_dynamical_2004}.
We then measure the power scaling characteristics of the ESUP process.
To verify the quadratic scaling of ESUP process with respect to
seed power, we vary it over almost five orders of magnitude and we
measure the generated idler power levels; they vary over eight orders
of magnitude. Starting from $350$$\mu W$ seed power in the bus waveguide
and successively reducing it to nanowatt level, the from the quadratically scaling idler
powers inferred effective conversion efficiencies (linear scaling) are plotted in Fig. \ref{fig:scaling}. From this we infer an on-chip normalized
$\eta_{ESUP}$ = 74,000 \%/W, in excellent agreement with the theoretically
predicted value.
\begin{center}
\begin{figure}
\begin{centering}
\includegraphics{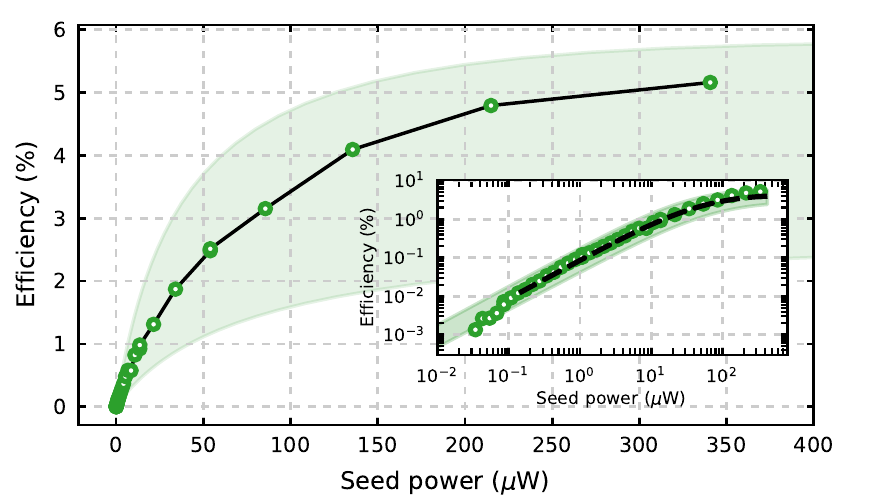}
\par\end{centering}
\caption{\label{fig:scaling} Idler generation as a function of seed power.
The main plot shows the inferred on-chip conversion efficiency as
a function of seed power with a clearly linear scaling (solid line) for the
unsaturated part. The inset shows a double-logarithmic plot of the whole measurement range of seed powers. The shaded region indicates the theoretical
expected trend within the estimated errors.}
\end{figure}
\par\end{center}

\begin{figure}
\begin{centering}
\includegraphics[width=0.95\columnwidth]{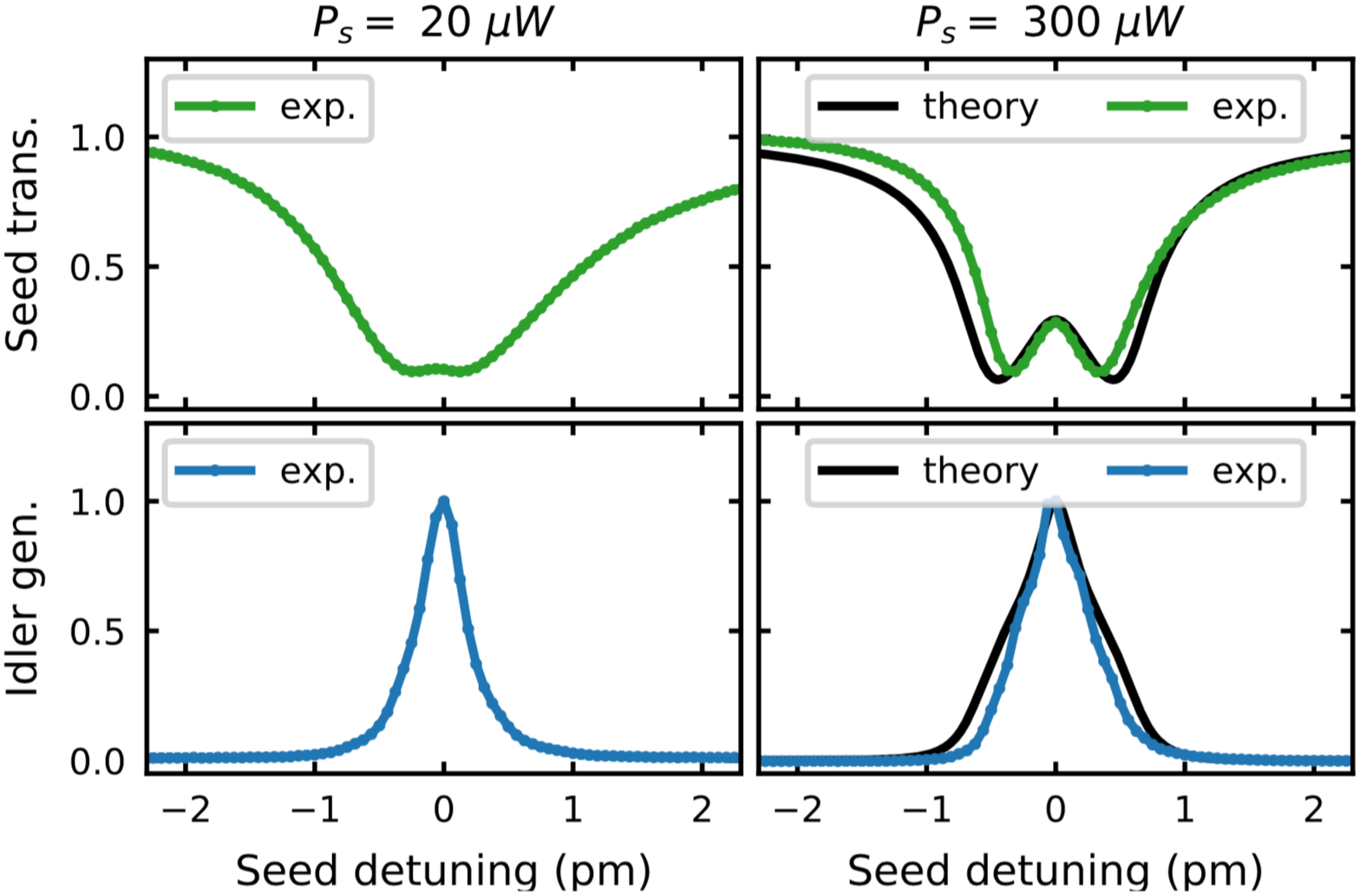}
\par\end{centering}
\caption{\label{fig:Signal-generation}Seed (green) transmission and idler
(blue) generation when seed wavelength is scanned across the resonance.
On the left, the seed  power $P_{seed}$= 20 $\mu W$. On the right,
where $P_{seed}=$ 300$\mu\text{W}$, the profiles are modified by
the strong nonlinear coupling.}
\end{figure}

We turn now to seed powers larger than a few tens of $\mu W$. Given
the lack of two-photon absorption, the saturation of the generated
idler observed in the main panel of Figure \ref{fig:scaling} suggests
that at such seed powers the nonlinear coupling between the modes
can no longer be treated perturbatively. To further investigate this
we scan the seed laser across the resonance wavelengths while the
idler power is monitored. Typical scans are shown in Fig. \ref{fig:Signal-generation}
for seed power $P_{seed}$= 50 $nW$ (left) and $P_{seed}=$ 300$\mu\text{W}$(right).
We see a gradual broadening of the seed peak as the seed power is
increased, until two dips can be clearly resolved. This is in excellent
agreement with the theoretical curves calculated starting from Eqs.
(1)-(2) (see Methods section) and using the nominal parameters
of the system without the need for any fitting. 

The splitting indicates a regime of ESUP in which the nonlinear coupling
rate $\Lambda_{3}|\beta_{P}\beta_{S}|$ between the modes is comparable
to the energy loss rate of the resonator, thus resulting in\textit{
strong nonlinear coupling}. Consequently a Rabi-like splitting of
the modes occurs. which is analogous to that theoretically investigated
by Carusotto and La Rocca \citep{carusotto_two-photon_1999}, who
focused on the second-order nonlinear interaction between two fields
oscillating at $\omega$ and $2\omega$ in a doubly resonant system.
They predict a Rabi-like resonance splitting at the fundamental frequency
when the nonlinear coupling rate exceeds the resonator damping rates. 

In the ring resonator, the steady state solutions for the mode amplitudes
predict the onset of this saturation as well as the presence of the
Rabi-like splitting of the resonance at $\omega_{S}$ (see Methods section). In particular, one can observe the splitting when the
``resolvability'' R, i.e., the ratio of the splitting to the line
width, is about unity. This quantity depends on the ring parameters
as well as the pump and seed powers according to 

\begin{equation}
R=\left|\frac{\Lambda\beta_{S}\beta_{P}}{\bar{\Gamma}_{S}}\right|\approx\frac{8\Lambda}{\omega_{S}}\sqrt{\frac{P_{P}P_{S}}{\hbar\omega_{S}^{2}\hbar\omega_{P}^{2}}}\frac{Q_{P}}{\sqrt{Q_{P}^{\mathrm{ext}}}}\frac{Q_{S}^{2}}{\sqrt{Q_{S}^{\mathrm{ext}}}},
\end{equation}
where $P_{P,S}$ are the pump and seed input powers and $Q_{P,S}^{\mathrm{ext}}$
are the quality factor of the resonators determined solely by the
external coupling. The last term of the equation is obtained by assuming
a linear solution for the pump and the seed powers and is strictly
valid for $R<1$, and yet it allows one to have a rough estimate of
when the strong nonlinear coupling regime can be achieved. 
\begin{center}
\begin{figure}
\begin{centering}
\includegraphics{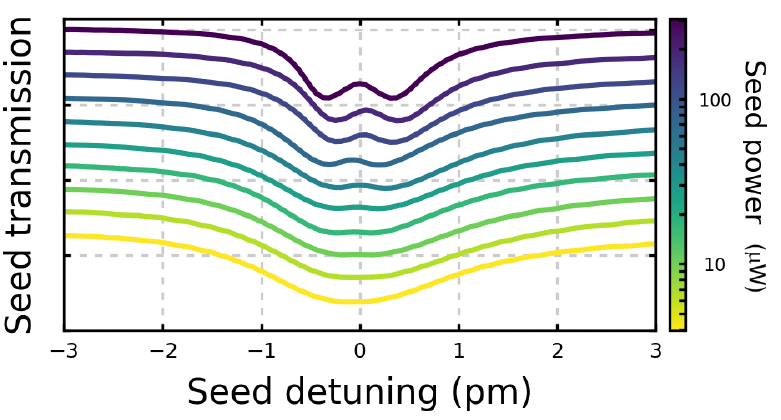}
\par\end{centering}
\caption{\label{fig:Seed-transmission}Seed transmission as the seed wavelength
is scanned across the seed resonance for different seed input power
levels. A characteristic change of the resonance shape appears for
higher seed power. }
\end{figure}
\par\end{center}

In Figure \ref{fig:Seed-transmission} we plot the seed transmission
as a function of detuning for several seed powers. It should be noticed
that unlike recent studies in which Rabi oscillations have been predicted
\citep{vernon_quantum_2016} and Rabi splittings have been experimentally
observed \citep{guo_-chip_2016} for frequency conversion in the strong
linear-coupling regime, here the strength of the coupling between
the fields at $\omega_{S}$ and $\omega_{I}$ \textit{depends on the
intensity of the seed field itself}, and in particular is linear in
$\beta_{S}$. Finally, we stress that this condition has been achieved
because our effective second-order nonlinear interaction is controlled
by the pump intensity, and occurs in a structure characterized by
an exceptionally large field enhancement and Q-factor. It is the combination
of these two elements that has allowed the first observation of a
nonlinear strong-coupling. 

\begin{figure}
\begin{centering}
\includegraphics{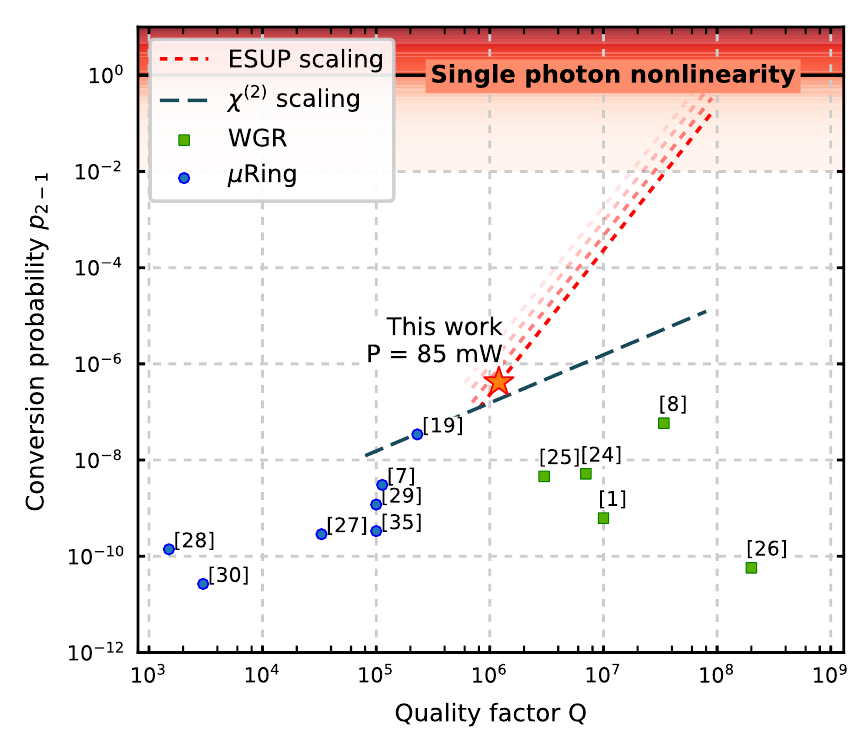}
\par\end{centering}
\caption{\label{fig:comparison}\textcolor{black}{Compari}son of two-photon
conversion probability $p_{2\rightarrow1}$ for different resonators
and material platforms: (green squares) whispering gallery mode resonators
(radius > 1 mm) \citep{ilchenko_nonlinear_2004,furst_naturally_2010,lin_wide-range_2013,lin_phase-matched_2016,lin_continuous_2014,furst_second-harmonic_2015},
(blue circles) microdisk resonators (radius < 100 um) \citep{kuo_second-harmonic_2014,mariani_second-harmonic_2014,lin_continuous_2014,lake_efficient_2016,xiong_integrated_2011,wang_second_2017,guo_aluminum_2017},
(star) this work. The probability is defined heuristically as $p_{2\rightarrow1}=\eta\frac{2E_{sp}}{\tau_{cavity}}=4\eta\frac{\hbar\omega^{2}}{Q}$
(where $2E_{sp}$is the energy of two photons to be converted and
$\tau_{cavity}$ the cavity dwelling time, and $\eta$ is the efficiency
of the corresponding parametric process)): indeed, for $p_{2\rightarrow1}$
approaching unity, quantization effects are expected to appear. Dashed
lines show the expected $p_{2\rightarrow1}$ as a function of Q and
on-chip pump power, (blue) for a $\chi^{(2)}$\textcolor{black}{{} (orange)
for ESUP. Importantly for }$\chi^{(2)}$devices\textcolor{black}{{}
$p_{2\rightarrow1}$} only scales\textcolor{black}{{} quadratically
with Q, whereas using ESUP the scaling with Q is cubic. Morover, the
red dotted line indicates the trend for a 100x improvement possible
without increasing Q. Together with a significant improvement in Q
up to 10s of Million as motivated by \citep{ji_ultra-low-loss_2017},
the extrapolation heuristically illustrates that approaching the $p_{2\rightarrow1}\simeq1$
regime is within reach of current $\mathrm{Si}{}_{3}\textrm{N}_{4}$
technology. }}
\end{figure}

The strength of this second-order nonlinear response raises the question
of whether the field associated with just a single pair of photons
at $\omega_{S}$ would be sufficient for a complete pair conversion
to a photon at $\omega_{I}$ in our system. In such a regime, new
phenomena are expected given the energy quantization of the optical
field \citep{chang_quantum_2014}. So far, much of the research has
been focused on atomic-like systems, either quantum dots or with atomic
clouds, while non-resonant photon-photon interactions in bulk nonlinear
materials have been far from approaching unit efficiency. For instance,
SHG driven by photon pairs has been demonstrated in LN waveguides,
but only yielding a very low probability $p_{2\rightarrow1}$ of a
two-photon conversion \citep{guerreiro_nonlinear_2014}. To further
increase $p_{2\rightarrow1}$ one can take advantage of light confinement
in micro resonators, either directly exploiting a natural second-order
material nonlinearity or a strong effective second-order response,
the latter of which we have done in this work. In Fig. \ref{fig:comparison}
we summarize several nonlinear resonator systems and show $p_{2\rightarrow1}$
as a function of the resonator quality factor $Q$, heuristically
defined as $p_{2\rightarrow1}=\eta\frac{2E_{sp}}{\tau_{cavity}}=4\eta\frac{\hbar\omega^{2}}{Q}$
(where $2E_{sp}$ is the energy of two photons to be converted, $\tau_{cavity}$
the cavity dwelling time, and $\eta$ is the efficiency of the corresponding
parametric process). We stress that one would expect quantization
effects to appear when $p_{2\rightarrow1}$ approaches a significant
fraction of unity. Currently, our system is characterized by $p_{2\rightarrow1}=$$4.7\times10^{-7}$
(star in Fig. \ref{fig:comparison}), which is the largest value ever
reported. This is despite the fact that our device has not yet been
optimized towards optimal coupling for all wavelength, with strong
over- and undercoupling for the pump and idler resonances, respectively.
Achieving critical coupling for all wavelength at the same intrinsic
Q and moderately increasing the pump power to 500mW would directly
yield an enhancement of $p_{2\rightarrow1}$ by 2 orders of magnitude.
Moreover, because $\eta$ scales with the 4th power of Q (see equation
\ref{eq:eta_sup}) it follows, that $p_{2\rightarrow1}$ scales with
$Q^{3}$. Thus a 30 times higher $Q$ would result heuristically already
in $p_{2\rightarrow1}\simeq1$.

This is challenging, for it requires bus-ring coupling optimization
at different wavelengths and the control and compensation of nonlinear
phase matching at higher pump powers, but these have individually
been shown already on the $\mathrm{Si}{}_{3}\textrm{N}_{4}$ platform,
and even more encouragingly $Q$ factors close to 70 million have
been recently obtained \citep{ji_ultra-low-loss_2017}. Based on these
simplified arguments we thus conclude that approaching the $p_{2\rightarrow1}\simeq1$
regime is within reach of current technology.

In conclusion, we demonstrated \emph{induced photon interactions}
based on effective second-order upconversion with a normalized efficiency
of 74,000 \%/W, and experimentally reached the regime of strong nonlinear
coupling for on-chip powers as low as few microwatts, in excellent
agreement with our theoretical model. The $\mathrm{Si}{}_{3}\textrm{N}_{4}$
platform is a mature and highly promising technology for obtaining
two-photon interaction at the single-pair level.

\section*{Methods}

To understand the dynamics of the nonlinear induced photon interaction
and ESUP process, we use a quantum mechanical Hamiltonian formalism
that fully captures the behavior of the system both in the few-photon
limit, and in the classical regime of strong coherent beams. A detailed
development of the formalism we employ can be found elsewhere in the
literature \citep{vernon_spontaneous_2015}; here we present a summary.

For the pump, signal, and idler, we identify intra-resonator annihilation
operators $b_{J}$ with $J\in\lbrace P,S,I\rbrace$ for the three
resonances that participate in the ESUP process. The Hamiltonian $H_{\mathrm{ring}}$
for the isolated resonant modes in the ring is then given by 
\begin{eqnarray}
H_{\mathrm{ring}}=\sum_{J}\hbar\omega_{J}b_{J}^{\dagger}b_{J}-\hbar\Lambda_{3}\left(b_{P}^{\dagger}b_{I}^{\dagger}b_{S}b_{S}+\mathrm{H.c.}\right),\label{ring_Hamiltonian}
\end{eqnarray}
in which we have retained only those nonlinear terms that participate
in the ESUP process. In particular, this Hamiltonian does not include
terms that represent self- and cross-phase modulation between the
three modes. However, for slowly-varying field envelope functions,
the effect of these processes is well-described by an effective shift
in the resonant frequencies $\omega_{J}$ \citep{vernon_strongly_2015}.
Since in this work we focus on CW input beams, this approximation
is well justified, and we assume the frequencies $\omega_{J}$ in
(\ref{ring_Hamiltonian}) already contain these corrections. The nonlinear
coupling constant $\Lambda_{3}$ is related to the third-order nonlinearity
of the waveguide material, and is estimated by 
\begin{eqnarray}
\Lambda_{3}\approx\frac{\hbar(\omega_{P}\omega_{S}^{2}\omega_{I})^{1/4}v_{g}^{2}\gamma_{NL}}{2\pi R},
\end{eqnarray}
where $v_{g}$ is the group velocity in the ring waveguide, $R$ the
ring radius, and $\gamma_{NL}$ the nonlinear parameter; for the ring
used in these experiments we find $\Lambda_{3}\approx10\mathrm{\;s}^{-1}$.

For each ring mode operator $b_{J}$ we also introduce a corresponding
input and output channel field operator $\psi_{J<}(t)$ and $\psi_{J>}(t)$
at the ring-channel coupling point, normalized such that $v_{J}\langle\psi_{J<}^{\dagger}(t)\psi_{J<}(t)\rangle$
($v_{g}\langle\psi_{J>}^{\dagger}(t)\psi_{J>}(t)\rangle$) is the
mean incoming (outgoing) photon flux in the side channel at frequencies
near $\omega_{J}$, with $v_{J}$ the corresponding group velocity.
The time dependencies of operators are understood in the usual Heisenberg
picture context. The coupling of these fields to the ring can be described
using standard cavity input-output theory \citep{gardiner_input_1985}.
Defining ring-channel coupling coefficients $\gamma_{J}$ to each
channel field/ring mode pair, the input-output relation for these
field operators is given by 
\begin{eqnarray}
\psi_{J>}(t)=\psi_{J<}(t)-\frac{i\gamma_{J}}{v_{J}}b_{J}(t),\label{input_output}
\end{eqnarray}
with $b_{J}(t)$ the time-dependent ring mode operator in the Heisenberg
picture.

The equations of motion in the classical limit for the ring mode amplitudes
can be found by calculating the full Heisenberg equations of motion
and replacing each operator by its expectation value. In what follows
we remove the fast optical time dependence and consider slowly-varying
quantities only. This gives 
\begin{eqnarray}
\frac{d\beta_{S}}{dt} & = & -\overline{\Gamma}_{S}\beta_{S}-i\gamma_{S}^{*}\alpha_{S}e^{-i\Delta_{S}t}+2i\Lambda_{2}(t)\beta_{S}^{*}\beta_{I}e^{i\Delta_{\mathrm{res}}t},\nonumber \\
\frac{d\beta_{I}}{dt} & = & -\overline{\Gamma}_{I}\beta_{I}+i\Lambda_{2}^{*}(t)\beta_{S}^{2}e^{-i\Delta_{\mathrm{res}}t}
\end{eqnarray}
for the seed and idler modes, where $\beta_{J}$ is the classical
ring mode amplitude for mode $J$, and $\overline{\Gamma}_{J}$ is
the full damping rate for mode $J$ (including both coupling to the
channel and to scattering from the resonator), related to the full
loaded quality factor $Q_{J}$ for that mode via $\overline{\Gamma}_{J}=\omega_{J}/(2Q_{J})$.
Here we have assumed that the seed is driven by a coherent beam with
amplitude $\alpha_{S}e^{-i\Delta_{S}t}$ in the channel, corresponding
to a seed input power $P_{S}=\hbar\omega_{S}v_{S}|\alpha_{S}|^{2}$
and detuning $\Delta_{S}$ from resonance. The detuning $\Delta_{\mathrm{res}}=2\omega_{S}-\omega_{T}-\omega_{P}$
accounts for any deviation of the resonances from perfectly equal
spacing in frequency. The effective second-order coupling coefficient
is defined as $\Lambda_{2}(t)=\Lambda_{3}\beta_{P}(t)$, with $\beta_{P}(t)$
the pump amplitude in the resonator.

In this work we assume a CW pump, so that the magnitude of $\Lambda_{2}$
is constant; The squared pump magnitude $|\beta_{P}|^{2}$ (corresponding
to the mean pump photon number in the resonator) in this regime is
given by 
\begin{eqnarray}
|\beta_{P}|^{2}=\frac{4P_{P}Q_{P}^{2}u_{P}}{\hbar\omega_{P}^{2}Q_{P}^{\mathrm{ext}}},
\end{eqnarray}
where $P_{P}$ is the input pump power in the channel, and $Q_{P}^{\mathrm{ext}}$
is the extrinsic quality factor of the pump associated with the channel-ring
coupling only, given by $Q_{P}^{\mathrm{ext}}=\omega_{P}v_{P}/|\gamma_{P}|^{2}$.
This expression assumes that the pump is sufficiently strong to render
any back-action or depletion effects on the pump mode negligible.
The degradation factor $u_{P}=(1+4l_{P}^{2})^{-1}$ ranges from 0
to 1, with $l_{P}$ the pump detuning from resonance in units of the
pump resonance full-width-half-max linewidth, $l_{P}=\Delta_{P}/(2\overline{\Gamma}_{P})$
with $\Delta_{P}$ the pump detuning from resonance. We define the
steady amplitude $\overline{\Lambda}_{2}$ via $\Lambda_{2}(t)=\overline{\Lambda}_{2}e^{-i\Delta_{P}t}$.

We seek a self-consistent stationary solution of Eqs. (\ref{eq:pw1},\ref{eq:pw2}).
Thus we search for a steady state solution of the form $\beta_{S}(t)=\overline{\beta}_{S}e^{-i\Delta_{S}t}$,
$\beta_{I}(t)=\overline{\beta}_{I}e^{-i\Delta_{I}t}$, with the overbar
denoting constant quantities. By substitution it directly follows
that this is only possible when the idler field oscillates with detuning
$\Delta_{I}=\Delta_{\mathrm{res}}-\Delta_{P}+2\Delta_{S}$; for what
remains we estimate $\Delta_{\mathrm{res}}=\Delta_{P}=0$ so $\Delta_{I}=2\Delta_{S}$.
The constant idler amplitude can then be written in terms of the seed
amplitude as 
\begin{eqnarray}
\overline{\beta}_{I}=\frac{i\overline{\Lambda}_{2}^{*}\overline{\beta}_{S}^{2}}{-i\Delta_{I}+\overline{\Gamma}_{I}}.\label{beta_I_eqn}
\end{eqnarray}
Inserting this back into the equation for the the seed, we obtain
\begin{eqnarray}
\left(-i\Delta_{S}+\overline{\Gamma}_{S}+\frac{2|\overline{\Lambda}_{2}|^{2}|\overline{\beta}_{S}|^{2}}{-i\Delta_{I}+\overline{\Gamma}_{I}}\right)\overline{\beta}_{S}=-i\gamma_{S}^{*}\alpha_{S}.\label{beta_S_eqn}
\end{eqnarray}
To solve for $\overline{\beta}_{S}$ we first need a value for $|\overline{\beta}_{S}|^{2}$.
This can be obtained by taking the modulus squared of (\ref{beta_S_eqn}),
which yields an implicit cubic equation for $|\overline{\beta}_{S}|^{2}\equiv N_{S}$:
\begin{eqnarray}
a_{3}N_{S}^{3}+a_{2}N_{S}^{2}+a_{1}N_{S}+a_{0}=0,\label{cubic_eqn}
\end{eqnarray}
where $a_{3}=4|\overline{\Lambda}_{2}|^{4}$, $a_{2}=4|\overline{\Lambda}_{2}|^{2}(\overline{\Gamma}_{S}\overline{\Gamma}_{I}-\Delta_{S}\Delta_{I})$,
$a_{1}=(\overline{\Gamma}_{S}\overline{\Gamma}_{I}-\Delta_{S}\Delta_{I})^{2}+(\Delta_{I}\overline{\Gamma}_{S}+\Delta_{S}\overline{\Gamma}_{I})^{2}$,
and $a_{0}=(\overline{\Gamma}_{I}^{2}+\Delta_{I}^{2})P_{S}/(\hbar Q_{S}^{\mathrm{ext}})$.
Note the explicit dependence of $a_{0}$ on the \emph{extrinsic} quality
factor for the seed resonance, making the coupling condition (under,
over, or critically coupled) of the seed a relevant parameter that
influences the conversion dynamics; indeed, the predictions of our
theory were found to depend quite sensitively on this coupling condition.

For a given set of ring parameters a positive real root $N_{S}$ of
Eq. (\ref{cubic_eqn}) can be found for each seed power and detuning,
representing a steady intra-resonator solution for the seed and idler
amplitudes. The seed amplitude can be calculated from (\ref{beta_S_eqn}),
and can be substituted in (\ref{beta_I_eqn}) to find the idler amplitude.
The input-output relations (\ref{input_output}) can then be used
to derive the output powers of the seed and the idler into the channel.
The transmission ratio of the seed is given by 

\begin{widetext}

\begin{eqnarray}
T=\bigg\vert1-2\Gamma_{S}\frac{-i\Delta_{I}+\overline{\Gamma}_{I}}{(-i\Delta_{S}+\overline{\Gamma}_{S})(-i\Delta_{I}+\overline{\Gamma}_{I})+2|\Lambda_{2}|^{2}N_{S}}\bigg\vert^{2},\label{seed_transmission}
\end{eqnarray}

\end{widetext}

and the corresponding generated idler power output to the channel
can be calculated from $P_{\mathrm{gen}}=2\hbar\omega_{I}\Gamma_{I}|\overline{\beta}_{I}|^{2}$.
In this expression and (\ref{seed_transmission}) the quantities $\Gamma_{S}$
and $\Gamma_{I}$ are the channel-ring coupling rates for the seed
and idler resonances, related to the corresponding extrinsic quality
factors via $\Gamma_{J}=\omega_{J}/(2Q_{J}^{\mathrm{ext}})$.

Viewed as a function of $\Delta_{S}$, the transmission ratio $T$
represents a seed transmission spectrum; this was used to produce
the theoretical curves in Fig. \ref{fig:Signal-generation} in the
main body. The corresponding generated power in the idler at each
point in these curves can be calculated, forming the theoretical spectral
curves for the generated idler in Fig. \ref{fig:Signal-generation},
as well as the power scaling in Fig. \ref{fig:scaling}.

In the limit of a weak seed input we can ignore the back-action on
the seed amplitude from its coupling to the idler, which enables a
simple analytic formula to be derived for the generated idler power.
Dropping the nonlinear term in the first of ((\ref{eq:pw1}) and solving
for $|\overline{\beta}_{I}|^{2}$, we find (after using the input-output
relation (\ref{input_output})) for the generated idler power output
in the channel 

\begin{widetext}
\begin{align*}
P_{\mathrm{gen}} & =\left(\frac{256\Lambda_{3}^{2}}{\hbar^{2}\omega_{S}^{4}\omega_{P}^{2}}\right)(P_{S})^{2}P_{P}(Q_{P}Q_{I}Q_{S}^{2})\frac{Q_{P}}{Q_{P}^{\mathrm{ext}}}\frac{Q_{I}}{Q_{I}^{\mathrm{ext}}}\left(\frac{Q_{S}}{Q_{S}^{\mathrm{ext}}}\right)^{2}u_{P}u_{\mathrm{net}},
\end{align*}

\end{widetext}

where $u_{\mathrm{net}}=(1+4l_{\mathrm{net}}^{2})^{-1}$ is the degradation
factor between 0 and 1 associated with the net detuning $l_{\mathrm{net}}$
(in units of idler resonance FWHM linewidth) associated with both
the pump and $\Delta_{\mathrm{res}}$. For our experiment both $u_{P}$
and $u_{\mathrm{net}}$ were estimated to be approximately unity,
corresponding to a perfectly resonant system. This expression for
the generated power demonstrates the quadratic scaling of the generated
idler power with the seed input, and can be used to predict the unit
power ESUP conversion efficiency $\eta_{ESUP}$ as defined in the
main body (eq. \ref{eq:eta_sup}), 
\[
\eta_{ESUP}=\frac{\mathrm{P_{\mathrm{gen}}}}{P_{S}^{2}}.
\]
Our formula takes into account different frequencies, coupling ratios
and quality factors for each resonance; the rough estimate of $\eta_{ESUP}$
in the main body is found by taking these all to be roughly equal
across the different resonances.

A few subtleties should be noted: first, the steady state solutions
described in this section for our experimental parameters are stable,
as verified by a first-order fluctuation analysis. However, unstable
regimes can be reached with realistic physical parameters, the presence
of which were found to sensitively depend on the seed coupling ratio.
Second, in order to obtain good agreement with the experimentally
measured values, it was necessary to include the effects of back-scattering
of the seed resonance in the resonator; this effect was strong enough
to be directly measured in our experiments. In particular, the presence
of back-scattering changes the procedure by which the coupling ratios
are extracted from measured linear transmission spectra of the microring
resonator. We modeled this phenomenologically with the inclusion of
a term $H_{\mathrm{BS}}=\hbar q(b_{S}b_{S-}^{\dagger}+\mathrm{H.c.})$
in the ring Hamiltonian, with $b_{S-}^{\dagger}$ the creation operater
for the counter-propagating mode at the seed frequency in the ring
that couples to the channel in the usual way, and where $q$ is a
backscattering strength parameter.

Both of these effects highlight the care which must be taken when
carrying out a mathematical analysis of nonlinear optics in high-Q
resonators, especially when considering strongly driven phenomena
like ESUP. 

Finally, to calculate the effective equivalent second-order nonlinearity
$\chi_{\mathrm{eff}}^{(2)}$ for our system, we study the nonlinear
part of the ring Hamiltonian that would describe a true degenerate
SHG/SPDC interaction, 
\begin{equation}
H_{\mathrm{SHG}}=-\hbar\Lambda_{\mathrm{SHG}}\left(b_{F}b_{F}b_{SH}^{\dagger}+\mathrm{H.c.}\right),\label{SHG_Hamiltonian}
\end{equation}
 where $F$ describes the fundamental and $SH$ the second harmonic
mode. The constant $\Lambda_{SH}$ describes the strength of the second-order
nonlinearity for this interaction, and is given by 
\begin{equation}
\Lambda_{\mathrm{SHG}}=\frac{\hbar^{1/2}\omega_{F}^{3/2}\chi^{(2)}}{2n^{3}\epsilon_{0}^{1/2}(LA)^{1/2},}\label{Lambda_SHG}
\end{equation}
 where $\omega_{F}$ is the frequency of the fundamental, $\chi^{(2)}$
the relevant component of the second-order nonlinear tensor, $n$
the material refractive index (assumed not to vary between the fundamental
and second harmonic frequencies), $L$ the ring length, and $A$ the
effective area.

The Hamiltonian (\ref{SHG_Hamiltonian}) is formally identical to
the Hamiltonian (\ref{Lambda_SHG}) that describes the ESUP process
in the case of a CW pump. Thus we can identify 
\[
|\Lambda_{3}\beta_{P}|=\Lambda_{\mathrm{SHG}}.
\]
 By calculating the pump amplitude $|\beta_{P}|$ and the coupling
strength $|\Lambda_{3}|$ in our experiment, using the the expression
(\ref{Lambda_SHG}) for $\Lambda_{\mathrm{SHG}}$, we can extract
the magnitude of $\chi^{(2)}$ that would be required in this system
to produce a comparable unit power conversion efficiency; this is
precisely the effective equivalent second-order nonlinearity $\chi_{\mathrm{eff}}^{(2)}$.
We find for a pump power of 85 mW, using $\Lambda_{3}=10$ s$^{-1}$,
that 
\[
\chi_{\mathrm{eff}}^{(2)}\approx0.6\mathrm{\;pm}/\mathrm{V}.
\]

\end{document}